\pdfoutput=1

\documentclass{svproc}

\usepackage[T1]{fontenc}
\usepackage[american]{babel}
\usepackage{graphicx}
\usepackage{multicol}

\usepackage{microtype} 

\usepackage{amsmath,amssymb,amsfonts}
\usepackage{mathtools}

\usepackage[backend=bibtex,url=false, doi=false,giveninits=true]{biblatex}
\usepackage{todonotes}
\usepackage{paralist}

\usepackage{graphicx}
\usepackage{textcomp}
\usepackage{xcolor}

\usepackage{url}
\usepackage{tikz}
\usepackage{microtype}
\usepackage{subcaption}

\usepackage{tabularx}
\usepackage{booktabs}
\usepackage{multirow}

\usepackage[ansinew]{inputenc}
\usepackage{cleveref} 
\let\cref\Cref
\let\citet\textcite

\begin{document}
\mainmatter

\title{Towards Explainable Scientific Venue Recommendations}
\titlerunning{Venue Recommendations}

 \author{Bastian Schaefermeier \and Gerd Stumme \and Tom Hanika}
\authorrunning{Schaefermeier et al.}


\institute{Knowledge and Data Engineering Group, University of Kassel\\
\email{\{schaefermeier, stumme, hanika\}@cs.uni-kassel.de}
}


\maketitle 

\begin{abstract}
  Selecting the best scientific venue (i.e., conference/journal) for
  the submission of a research article constitutes a multifaceted
  challenge. Important aspects to consider are the suitability of
  research topics, a venue's prestige, and the probability of
  acceptance. The selection problem is exacerbated through the
  continuous emergence of additional venues. Previously proposed
  approaches for supporting authors in this process rely on 
  complex recommender systems, e.g., based on Word2Vec or
  TextCNN. These, however, often elude an explanation for their
  recommendations. In this work, we propose an unsophisticated method
  that advances the state-of-the-art in two aspects: First, we enhance
  the interpretability of recommendations through non-negative matrix
  factorization based topic models; Second, we surprisingly can obtain
  competitive recommendation performance while using simpler learning
  methods.  
\end{abstract}

\section{Introduction}
An essential part of the scientific research process is the
publication of the obtained results at a suitable venue, i.e., a
particular conference, workshop, or journal. The related selection
problem for the best fitting scientific venue has many different
aspects, such as the fit of the research topics, the prospects of
acceptance, and the prestige of the venue. The complexity of the
selection is further exacerbated by the growing number of publication
venues, the increasing granularity of research topics, and the
exponentially surging number of publications.

To support researchers with this task, different methods have been
proposed, e.g., based on Latent Dirichlet
Allocation~\cite{venuerecabstract}, hybrid approaches incorporating
social networks~\cite{cnaver, hybridrec}, or procedures that draw from
background ontologies~\cite{sbr, cso}. Moreover, recent approaches
based on deep learning methods achieved high accuracy in
recommendations~\cite{wts}. All these methods have in common that
their recommendations are insufficiently explained. For example,
\textcite{wts} solely highlight words from the input article that were
essential for a recommendation.

With the present work we show a new approach for recommending venues
that improves on explainability. From the information a scientist
provides, such as paper title, abstract and, possibly, a list of
keywords, our method creates a ranking over $k$ thematically fitting
venues. 
%
Our method further generates a list of the most important research
topics found in the input article. We provide a contrasting
juxtaposition to the venues by generating a similar list of research
topics for each recommended venue. The research topics are found
automatically through a topic model, specifically a non-negative
matrix factorization (NMF), that is precomputed on a corpus of
suitable research articles.
The thereof resulting recommendations entail two major
advantages. First, they are more comprehensible due to their annotations
using research topics. Second, their explanations go beyond the mere
occurrence of certain terms in the input article.

From a procedural point of view, through NMF, we identify research topics in a
corpus and represent venues as a combination of these topics. This
approach has been used successfully before to map trajectories of
venues within thematic spaces~\cite{tps} in an interpretable manner.
Based on that our specific contribution with the present work is
threefold.
\begin{inparadesc}
\item[First,] we present a simple yet effective recommendation
  procedure based on tf-idf, logistic regression and non-negative
  matrix factorization.
\item[Second,] we demonstrate that our setup can achieve competitive
  or even better results with respect to the state-of-the-art in venue
  recommendation. We base this claim on the comparability accomplished by
  using the data from~\citet{wts}.
\item[Third,] we show how the human interpretability of venue
  recommendations may be enhanced using research topics discovered by
  NMF.

  
\end{inparadesc}
\section{Problem Description}
When researchers are looking for a publishing venue for their recent
research work $d$, they encounter an increasing number of conferences and
journals. Each of said venues $v\in V$ has properties $p\in P$, such as prestige,
acceptance rate, and emphasized research topics, to be considered for
submission.
\begin{problem}[General Submission Problem] \ \\
  Given paper (document) $d$, generate a relevance ranking over
  a set $V$ of venues.
\end{problem}

The order of this ranking should reflect the similarity between
elements of $V$ and the input paper (document) $d$ in terms of their
respective research topics. Furthermore, other properties of $P$ may
be reflected with different levels of importance. The resulting
top-$k$ ranked venues can be viewed as individual venue
recommendations for paper $d$.  In order to cover the thematic aspect
of the submission problem we resort to a paper corpus $C$ over a
research domain. This corpus is constituted by a set of papers labeled
with their publishing venues $v \in V$. In the following, we consider
the aspect of finding good topical fits as the submission problem. We
require that all venues $V$ are represented by the papers in $C$,
i.e., for any $v$ there is a sufficient number of papers in
$C$. Moreover, the topics of $d$ must be from the same particular
research domain as the corpus $C$.

\citet{betterscience} pointed out that transparency is a necessary
feature of search engines for research. The most prominent reasons are
accountability of the search results and the disclosure of possible
biases (e.g., in algorithms or data). In our work, we hence do rather
not focus on outperforming previous benchmarks, but on advancing
methods that are intrinsically comprehensible. Hence, our approach is
based on a \emph{topic model} that is derived from non-negative matrix
factorization (NMF). This method has been used in previous works to
gain explainable insights into research topics \cite{tps}. Using NMF
we annotate venues, i.e., conferences and journals, with their
respective topics. Furthermore, we attempt to use the discovered topics
as an input to the ranking method. What is more, our method, as
presented in this work, is useful to find similar research for a given
piece of work, although our focus here is with the first application
in mind.

\section{Related Work}
\citet{venuerecstyle} recommend venues from ACM digital library using a collaborative filtering approach incorporating topics and writing style. \citet{venuerecabstract} test three different methods based on n-grams and Latent Dirichlet Allocation. More complex, hybrid methods integrate social network analysis \cite{cnaver, hybridrec}. All of the above do not add explanations to recommendations.

Smart Book Recommender (SBR) represents books, journals and conference proceedings as vectors of topic counts extracted through the \emph{Computer Science Ontology} (CSO) \cite{sbr, cso}. 
Due to the dependence on CSO, SBR is not directly applicable to other research domains. Recommendations are only available for items in a pre-computed database. 
%
\citet{ianaetal} develop a conference recommender based on SciGraph, a taxonomy by Springer Nature. The authors experiment with author information, paper abstracts and paper keywords. For abstracts, different representations are evaluated, such as TF-IDF, Latent Semantic Analysis (LSA), probabilistic LSA and different embedding methods (Word2Vec \cite{word2vec}, GloVe and FastText). For keywords, SciGraph marketing codes are used, categories defined by Springer. The best performing method is based on these marketing codes, which, however, may not be available from other than the used data sources.

In ``Where To Submit'' (WTS, \citet{wts}) self-trained Word2Vec embeddings and TextCNN \cite{textcnn} are employed. Enhancements are made by incorporating paper titles and keywords and applying convolutions separately to them. 
Explanations through the
\emph{integrated gradients method} are added \cite{integratedgradients}. 
This computes words influential for the classifier, which are visually
highlighted. These, however, do not necessarily improve the
interpretability. For example, often, only few words are highlighted.
Examples for non-interpretable black-box recommenders can be found online.\footnote{\url{https://journalsuggester.springer.com}} These are often restricted to journals or specific publishers.

\section{Method}
\label{sec:method}
We approach the submission problem as a supervised learning task. For every paper $c$ in a corpus $C$, we use its publication venue $v \in V$ as its ground truth label. Based on that we train a supervised classifier to predict the venue. 
The output of such classifiers are probabilities for, or fractions of, class membership, which impose a linear order on $V$.

Our aim is to generate both, recommendations and explanations comprised of
interpretable, automatically derived research topics. Our reasoning is that research topics
provide more informative explanations for venue recommendations than
single words \cite{tps}. In short, the envisioned NMF approach
allows for incorporating words that are not part of the input document $d$ (i.e., the paper title, abstract and keywords) into the explanation.
This additional information is pre-extracted from co-occurrences in a
paper corpus of the research domain. The research topics allow for a unified view on venues as well as documents.

For the rest of this work, a document is comprised of the following features: a title; an abstract; a set of keywords. Each of these attributes is a string (or set of strings). Every document is labeled with its publication venue $v \in V$.
\begin{definition}[Paper Corpus]
A \emph{paper corpus} is a set $C \subseteq T \times A \times 2^{S} \times V $. A (research) paper $c\coloneqq (t,a,s,v) \in C$ consists of a title $t \in T$, an abstract $a \in A$, a set of keywords $s \subseteq S$ and is labeled with a venue name $v \in V$.
\end{definition}

\subsection{Non-negative Matrix Factorization}
We use non-negative matrix factorization \cite{nmf} to create a topic
model from a corpus of papers and to obtain a lower-dimensional
representation of the papers in topic space. It has previously been
shown that NMF is able to reconstruct research topics that are similar
to research categories produced by human experts \cite{tps}. For an
input matrix $U \in \mathbb{R}_{\ge 0}^{u\times n}$ containing
document representations (e.g., as tf-idf vectors), NMF finds an
approximate factorization $U \approx WH$. The output matrices $W \in
\mathbb{R}_{\ge 0}^{u\times l}$ and $H \in \mathbb{R}_{\ge 0}^{l
  \times n}$ contain a set of basis vectors as well as the desired
lower-dimensional representation of the input documents (represented
as a weighted sum of the basis vectors). The $l$ basis vectors can be
interpreted as vectors of term importances, i.e. \emph{topics}. The
lower-dimensional document representations can be interpreted as
proportions of topics in a document.  We may stress that NMF, due to
its additive nature, allows for human interpretable topics as well as
document representations as vectors. We will use research topics
calculated through NMF as an attempt towards recommendation
explanations.

\subsection{Classification Methods}
The proposed approach for creating explanations is independent from
the concrete choice of the classification method.  Hence, we are able
to evaluate a multitude of procedures. We also evaluate different
methods for feature extraction and paper representation. In
particular, we use \emph{tf-idf} vectors \cite{tfidf} as well as the
aforementioned NMF topics as features.  This being said, we emphasize
in our investigation the \emph{logistic regression} approach, since
this has already been successful in previous work \cite{wts,
  ianaetal}. Additionally, logistic regression is a robust procedure
that does not require a large number of parameter values to be
identified. Furthermore, the resulting classification function can be
computed fast, unlike e.g., similarity-based approaches
\cite{sbr}. Ultimately, linear models such as logistic regression are
already interpretable to some extent \cite{Burkart_2021}. In contrast
to \cite{wts}, we prefer tf-idf over tf (i,e., term frequencies), because
we assume, as being one de-facto standard in text representation, it
may enable better classification results. In our experiments we
compare our results to~\citet{wts}.

We use the following methods:
\begin{inparadesc}
\item[$\bullet$~Uniform Random~--]
we predict venues in random order.
\item[$\bullet$~Most Frequent~--]
we predict venues in the order of their prevalence in the data set.
\item[$\bullet$~Logistic Regression (``Logit'')~--]
we employ multi-class logistic regression from \emph{scikit-learn} \cite{logit, sklearn}.
\end{inparadesc}
We apply these classifiers to different document representations. For every $c\in C$, we concatenate the title, abstract and
keywords before calculating the representations:
\begin{inparadesc}
\item[$\bullet$~NMF~--]
we use the representation calculated through NMF~\cite{nmf} using a
reasonable topic number. We adapt the choice for \texttt{kappa} (learning rate) and \texttt{w\_max\_iter} and \texttt{h\_max\_iter}, the maximum training iterations per batch for matrices $W$ and $H$. We report their values in \cref{sec:experiments}. 
\item[$\bullet$~tf-idf~--]
we use term-frequency (tf) as well as tf-idf \cite{tfidf}, i.e., the product of tf and inverted document frequency (idf).
\item[$\bullet$~NMF~+~tf-idf~--]
we concatenate both of the above representations to a single vector.
\end{inparadesc}

\section{Experimental Evaluation}\label{sec:experiments}
We evaluate our methods on two different domains: Artificial
intelligence (AI) and medicine (MED). More specifically, we use the
data sets\footnote{\url{https://github.com/konstantinkobs/wts}}
compiled and described by \citet{wts}, which are based on Semantic
Scholar \cite{semanticscholar}. Each data set is comprised of papers
from 78 non-uniformly distributed classes (i.e., venues). The AI
corpus contains 245,573 papers; the MED corpus 2,924,609 papers.

To facilitate comparability we calculate the $accuracy$, the $accuracy@k$ and the mean reciprocal rank (MRR) on a test data set of papers. For $accuracy@k$, a ranking for a tested paper is counted as correct, when its true label is contained in the top $k$ ranked venues. This count is divided by the total size of the test set. 
\citet{ianaetal} used the term $recall@k$, which, due to the single ground truth label, calculates the same.
MRR is calculated from a test set $\mathbb{T} \subset C$ as follows: Let $rank(x, \mathbb{C}) \in \mathbb{N}_{> 0}$ be the rank of the true label of publication $x \in \mathbb{T}$ in a ranking generated by algorithm $\mathbb{C}$. Then MRR is the average of the inverse ranks, i.e., $\frac{1}{|\mathbb{T}|} \sum_{x \in \mathbb{T}}{\frac{1}{rank(x, \mathbb{C})}}$.
For NMF, we use 100 topics for AI and 500 topics for MED to account
for its larger size. We set $\texttt{kappa}$ to $1$ and the
values of \texttt{w\_max\_iter} to 300 and \texttt{h\_max\_iter} to
100.
This led to better training convergence.

\subsection{Results and Discussion}

\begin{table}
  \caption{\emph{Left}: Results of different recommendation methods. Results
    for WTS are taken from~\citet{wts}. \emph{Right}: Top 3 recommendations
    and top 3 topics for the BERT paper \cite{bert}. Our
    recommendation NAACL is the true publication
    venue.}\label{tab:scores}
\resizebox{0.49\columnwidth}{!}{
\begin{tabular}{ll|lll}
\toprule
& \textbf{Method} & \textbf{Acc} &\textbf{Acc@5} & \textbf{MRR}\tabularnewline
\midrule
\multirow{7}{*}{\rotatebox{90}{\bfseries AI}} 
& Uniform Random (\emph{avg}) & 0.012 & 0.067 & 0.063 \\
& Most Frequent & 0.086 & 0.319 & 0.212 \\
& WTS &0.503 & 0.831 & 0.645 \\
&Logit (NMF) & 0.397 & 0.744 & 0.550 \\
& Logit (tf) & 0.464 & 0.787 & 0.605 \\
&Logit (tf-idf) & 0.509 & 0.841 & 0.651 \\
&Logit (tf-idf + NMF) & \textbf{0.510} & \textbf{0.843} & \textbf{0.652} \tabularnewline
\midrule
\multirow{6}{*}{\rotatebox{90}{\bfseries MED}} 
& Uniform Random (\emph{avg}) & 0.013 & 0.064 & 0.063 \\
& Most Frequent & 0.069 & 0.213 & 0.157 \\
& WTS & \textbf{0.659} &  \textbf{0.948} & \textbf{0.782} \\
&Logit (NMF) & 0.365 & 0.762 & 0.538 \\
& Logit (tf-idf) & 0.591 & 0.917 & 0.729\\
& Logit (tf-idf + NMF) & 0.592 & 0.917 & 0.730 \\ 
\bottomrule
\end{tabular}
}
\resizebox{0.499\columnwidth}{!}{
\begin{tabular} {p{8.55cm}}
\toprule
\textbf{Input Paper}\\
\textbf{Title:}
\textit{BERT: Pre-training of Deep Bidirectional ($\ldots$)} \\
\textbf{Abstract:}
\textit{We introduce a new language ($\ldots$)}\\
\textbf{Keywords:} <None>\tabularnewline
\midrule
\textbf{Found Topics (in rank order)}
  \begin{compactenum}[a)]
  \item question, answering, questions, answer, answers
  \item language, natural, processing, parsing, dialog
  \item deep, convolutional, neural, network, cnn
  \end{compactenum}\ \vspace{-0.45cm} \\ 
\midrule
\textbf{Recommendations (in rank order)}
  \begin{compactenum}[1)]
  \item \emph{EMNLP}:\hspace{0.05cm}
    \begin{minipage}[t]{0.9\linewidth}
    \begin{compactenum}[a)]
  \item translation, machine, statistical, bleu
  \item parsing, grammar, parser, dependency
  \item corpus, text, corpora, news
  \end{compactenum}
    \end{minipage}
  \item \emph{NAACL}:\hspace{0.05cm}
    \begin{minipage}[t]{0.9\linewidth}
  \begin{compactenum}[a)]
  \item translation, machine, statistical, bleu
  \item corpus, text, corpora, news
  \item word, dictionary, sense, disambiguation
  \end{compactenum}
    \end{minipage}
\item \emph{AAAI}:\hspace{0.05cm}
  \begin{minipage}[t]{0.9\linewidth}
  \begin{compactenum}[a)]
  \item reasoning, representation, knowledge, case
  \item optimization, problem, constraint, constraints
  \item agent, agents, multi, multiagent
  \end{compactenum}
  \end{minipage}
\end{compactenum}\ \vspace{-0.2cm} \\ 
\bottomrule
\end{tabular}
}
\end{table}

\cref{tab:scores}~(left) depicts our scores in comparison to the
state-of-the-art WTS. Notably, on AI, we find that logistic regression
achieves higher values for all performance measures. This is
remarkable, given the simplicity of logistic regression and the
complexity of WTS. \citet{wts} also evaluated logistic regression,
however, using mere term frequencies instead of tf-idf. We hence added
an experiment on AI to confirm that tf-idf is favorable. On the 
larger MED corpus, WTS exhibits the best performance scores. This shows
that the employed TextCNN and Word2Vec profit from large amounts of
training data, a fact that has been frequently stated for
neural networks before.  MED is a very diverse data set covering
venues from the fields chemistry, medicine, physics and more. In
practice, the domain-focussed AI corpus is a more realistic
recommendation scenario.

Adding NMF-topics to tf-idf vectors led to slightly higher scores for
AI and MED. The effect might be larger when NMF is trained on an
additional corpus or when NMF variants incorporating class information
are used~\cite{supervisednmf}.
Pure NMF representations led to lower scores than tf-idf. Yet, despite
their few vector components they are in the range of the tf
representations.
To a certain degree, NMF topics allow for an interpretation and
assessment of recommendations.  \cref{tab:scores} (right) depicts
recommendations and topics for an example paper $d$~\cite{bert}. The true
venue is ranked second. The identified topics of $d$,
represented by their top-weighted terms, are interpretable as
\emph{question answering}, \emph{natural language processing} and
\emph{neural networks}.
The venue topics are often strongly related and, where not, may give a
hint for further recommendation assessment.

\section{Conclusion and Outlook}
We presented methods towards explainable scientific venue
recommendations.\footnote{Demonstration available at
  \url{https://sci-rec.org}} First, we showed that logistic regression
with tf-idf is a competitive recommendation setup. Second, we
illustrated a principled approach for annotating recommendations with
topics derived from a research corpus. The so-provided contrasting
juxtaposition using topics allows for explanations that exceed the
content of the input paper (i.e., query). We envision that future work
should target automated topic labeling \cite{topiclabels} to further
increase topic comprehensibility.


\printbibliography

\end{document}